\documentclass[unnumsec,webpdf,contemporary,large,anonymous]{oup-authoring-template}%

\graphicspath{{Fig/}}

\theoremstyle{thmstyleone}%
\theoremstyle{thmstyletwo}%
\theoremstyle{thmstylethree}%

\begin{document}

\journaltitle{Behaviour and Information Technology}
\DOI{DOI added during production}
\copyrightyear{YEAR}
\pubyear{YEAR}
\vol{XX}
\issue{x}
\access{Published: Date added during production}
\appnotes{Paper}

\firstpage{1}


\title[Temporal Decoupling Pointing System for External Objects]{Designing and Evaluating In-Vehicle Temporal Decoupling Pointing System for Selecting External Object}

\author[1]{Jaehoon Pyun}
\author[1]{Younggeol Cho}
\author[1]{Seon Gyeom Kim}
\author[1]{Woohun Lee}
\author[2,$\ast$]{Donghyeon Ko}

\address[1]{KAIST, Daejeon, South Korea}
\address[2]{University of Ulsan, Ulsan,South Korea}

\corresp[$\ast$]{Corresponding author. \href{mailto:donghyeonko@ulsan.ac.kr}{donghyeonko@ulsan.ac.kr}}


\received{Date}{0}{Year}
\revised{Date}{0}{Year}
\accepted{Date}{0}{Year}

\abstract{As In-Vehicle Infotainment Systems (IVIS) grow in complexity, selecting external points of interest (POIs) using traditional touchscreens significantly increases driver cognitive load. Recent evidence indicates that this visual-motor overload induces dangerous "hand-before-eye" behaviors, degrading primary driving tasks. To address this, we propose Point and Select, a novel in-vehicle interaction paradigm that introduces temporal decoupling to spatial gestures. By dividing the interaction into a rapid, ballistic spatial anchoring phase ("Rough Pointing") and a deferred, tactile confirmation phase ("Fine Selection"), our design aligns with the driver's cognitive-motor sequence. We evaluated this temporally decoupled approach in a high-fidelity driving simulator under urban speed conditions. Results indicate that Point and Select effectively minimizes perceived cognitive workload while seamlessly maintaining primary driving performance. This study demonstrates that decoupling spatial identification from confirmation successfully mitigates cognitive friction, offering a safer behavioral design strategy for non-autonomous driving environments.}

\maketitle

\section{Introduction} \label{sec:intro}

With the advances of automotive technologies, In-Vehicle Infotainment Systems have become a focal point of modern vehicles, incorporating real-time information, entertainment, and communication services \cite{chen2019influence, lee2014dynamics, pfleging2015driving}. Among its functionalities, route management enables drivers to efficiently select destinations and points of interest during their journeys. While driving, drivers frequently make spontaneous decisions to change routes or explore nearby locations. Yet in many cases, the driver is unaware of the specific name or address of a POI visible through the windshield—a building, restaurant, or landmark that catches the eye—and traditional navigation systems offer no direct means to address this.

Providing input from external surroundings to the IVIS could unlock powerful new scenarios: requesting guidance to a building's parking entrance, saving a visible restaurant for later, or identifying an unknown structure \cite{cohen2016hri, sauras2017voge, tscharn2017stop}. Currently, drivers resort to manual phone searches, noting place names, or interacting directly with the navigation display. These methods are not only time-consuming but pose significant safety risks by diverting attention, hands, and cognitive focus from the primary task of driving \cite{blaschke2009driver, eysenck2001principles, naujoks2016secondary, wierwille1993demands}. Recent evidence highlights that in-vehicle touchscreens induce a dangerous ``hand-before-eye'' distraction \cite{shen2025touchscreens}, a concern now reflected in the 2026 Euro NCAP protocols, which penalise screen-centric operations and mandate physical controls for essential tasks \cite{euro_ncap_2026}. While autonomous‑driving advances shrink the driver’s active control role such as Tesla Full Self‑Driving and GM Super Cruise, the ongoing Society of Automotive Engineers Level (SAE) Level 2–3 transition still demands that they remain ready to take over at any moment, making low‑effort, low‑distraction designs essential for safety. Even with sufficient attention, dynamic urban elements such as crossing pedestrians drastically reduce the safe dwell time available for interacting with external targets \cite{huang2025gaze, merlino2019crossing, pomarjanschi2012gaze}.

Underscoring the need for interaction techniques considering speed and safety, spatial interaction using natural gestures like finger, pointing to reference objects in the environment, stands out as a promising approach which is an intuitive, everyday behaviour \cite{droeschel2011learning, fransen2007using}, and prior work has demonstrated its potential for selecting external objects from within a vehicle \cite{gomaa2020studying, weidner2019interact, graichen2019evaluation}. However, these investigations have predominantly targeted highly automated vehicles (SAE Level 3--4), where the driver is largely disengaged from the road, and have focused almost exclusively on selection accuracy without evaluating the concurrent impact on primary driving performance. The gap that remains is a technique suited to the SAE Level 2--3 transitional period, where drivers are still actively engaged and distraction costs are highest. Moreover, introduces not only biomechanical instability but also severe cognitive friction, further motivating a fundamentally different design\cite{mayer2018effect, colley2021swivr}.

In this paper, we explored an interaction technique called \textbf{Point and Select} for inputting external POIs into the IVIS during the SAE Level 2--3 transitional period. Grounded in empirical observations of how drivers naturally reference their surroundings, Point and Select employs a two-phase process: a \textit{Rough Pointing} phase, in which the driver rapidly and approximately indicates a target via finger gesture at the moment it appears; and a \textit{Fine Selection} phase, in which the driver confirms or adjusts the candidate using a steering-wheel button through temporal decoupling, allowing drivers to capture POIs instantly while safely delaying the cognitive load of confirmation until driving demands decrease, thereby maintaining primary driving performance.

\vspace{0.2cm} \noindent The primary contributions of this work are threefold: 

\begin{itemize} 
\item A video ethnography revealing that drivers point roughly and delay confirmation until driving demand drops, motivating a temporally decoupled design.
\item \textit{Point and Select}, a two-phase interaction that temporally decouples rough finger pointing from fine selection via steering-wheel backward navigation.
\item A simulator evaluation under manual (SAE Level 2) conditions, empirically demonstrating how temporal decoupling minimizes perceived cognitive workload while maintaining primary driving performance.
\end{itemize}

In the remainder of the paper, we report on a video ethnography that motivated our design, followed by the conceptualization and implementation of the \textit{Point and Select} prototype within a simulated driving environment. To evaluate our approach, we detail a user study, present its results, and discuss the implications of our findings before concluding the paper.

\section{Related Work} \label{sec:related} %

\subsection{Interactions for Secondary Driving Tasks}

Prior work has investigated a range of interaction modalities for secondary in-vehicle tasks, each revealing distinct trade-offs. Touch-based interfaces remain dominant in consumer vehicles, yet empirical studies report that their visual cost is compounded by icon size and display density on instrument clusters \cite{pfleging2012multimodal, sun2024optimizing}. As task complexity grows, drivers begin reaching for the screen before they have visually located their target, increasing lane deviation \cite{shen2025touchscreens}. Voice-based interfaces reduce manual demand but substitute it with a memory retrieval burden. Drivers need to recall precise commands or place names and recognition accuracy drops sharply in noisy cabin environments \cite{sauras2017voge, jung2021effect}. Combining voice input with a touchpad partially mitigates this friction \cite{jung2020voice+}, yet these approaches are not able to refer to an external object whose name or address the driver does not know. Furthermore, in urban settings, vulnerable road users and unpredictable events compress the window for secondary interaction to only a few seconds \cite{pomarjanschi2012gaze}, and drivers naturally interleave subtasks into brief, self-paced gaps between driving demands \cite{naujoks2017importance, brumby2009focus}.

Since Bolt \cite{bolt1980put} demonstrated that deictic gestures combined with speech can directly reference on-screen objects, spatial pointing has been widely adopted in Human-Computer Interaction \cite{droeschel2011learning, folmer2012spatial, fransen2007using}. In the driving context, Rümelin et al.\ \cite{rumelin2013free} showed that free-hand pointing enables drivers to indicate distant roadside objects without diverting gaze to a display. Subsequent work has improved spatial interaction in vehicles through various strategies: enhancing unimodal precision \cite{weidner2019interact, graichen2019evaluation}, fusing pointing with gaze to relax accuracy requirements \cite{gomaa2020studying, jia2018gaze, lakier2019cross, gazepointAR2024}, and incorporating auditory feedback to reduce visual confirmation demands \cite{sterkenburg2019design, cao2024head, martinez2017}. Despite their diversity, however, all of these approaches share a common structural constraint: the driver must indicate \textit{and} confirm the target while it is still in view and the gesture is actively held—a \textit{synchronous coupling} of spatial capture and selection confirmation.

In driving, however, the window in which a target is visible and the window in which the driver can spare attention for confirmation rarely overlap. Moreover, discrete mid-air confirmation gestures (e.g., a pinch or dwell) destabilise the pointing vector through vehicle vibration \cite{mayer2018effect, colley2021swivr}, so the moment of best spatial opportunity coincides with the highest selection instability and cognitive load. As Table~\ref{tab:related} summarises, this constraint has limited prior evaluations to SAE Level~3--4 automation, where drivers are relieved of vehicle control, and the concurrent effect on primary driving performance remains underexplored.

\begin{table*}[t] 
\centering
\caption{Comparison of related work on spatial interaction in driving contexts. All prior studies assume SAE Level 3--4; none jointly evaluates selection performance and primary driving degradation.} 
\label{tab:related} 
\begin{tabular}{p{2.8cm}p{3.0cm}p{2.4cm}p{3.0cm}p{3.0cm}}\hline 
\textbf{Study} & \textbf{Modality} & \textbf{Driving Level} & \textbf{Evaluation Metrics} & \textbf{Safety / CL Analysis} \\ 
\hline 
Gomaa et al. \cite{gomaa2020studying} & Finger + Gaze & L4 (Autonomous) & Accuracy & Not reported \\ 
Tscharn et al. \cite{tscharn2017stop} & Gesture + Voice & L3--4 & Usability & Limited \\ 
Sterkenburg et al. \cite{sterkenburg2019design} & Spatial + Audio & L4 & SUS, Usability & Partial \\ 
Cao et al. \cite{cao2024head} & Head Gesture & L3--4 & SR, Usability & Not reported \\ 
Weidner et al. \cite{weidner2019interact} & Multi-modal & L4 & SR, TCT, DALI & Not reported \\ 
\textbf{Point and Select (Ours)} & \textbf{Finger + Steering-wheel Button} & \textbf{L2 (Non-autonomous)} & \textbf{SR, TCT, LLM, SM, DALI, NASA-TLX} & \textbf{Road-regulation-based analysis} \\ 
\hline 
\end{tabular} 
\end{table*}

\subsection{Temporal Decoupling and Coarse-to-Fine Selection} \label{sec:temporal_decoupling}
Zhai et al.\ \cite{zhai1999magic} and Cockburn et al.\ \cite{cockburn2009review} explored \textit{coarse-to-fine} interaction, in which a rapid, approximate first action narrows the candidate space and a deliberate second action confirms the choice. Extending this principle, several studies have shown that the two phases need not occur simultaneously: separating them in time—\textit{temporal decoupling}—allows a user to capture a spatial reference at the moment of opportunity and defer confirmation to a self-paced later moment \cite{blanco2006impact, eysenck2001principles}. In vehicle ergonomics, Lee et al.\ \cite{lee2015wheel} demonstrated that relocating discrete input to a steering-wheel button keeps the driver's hands anchored and reduces postural interference. Together, these findings suggest that combining temporal decoupling with an on-wheel confirmation could address the synchronous-coupling limitation identified above, yet this combination has not been applied to in-vehicle spatial selection, particularly under SAE Level~2 conditions.

\subsection{Dual-Task Measurement in Driving}

Evaluating secondary in-vehicle tasks requires instruments sensitive to both subjective workload and objective driving degradation. Pauzié et al.\ \cite{pauzie2008method} developed the Driver Activity Load Index (DALI), an adaptation of the NASA Task Load Index \cite{hart1988development} calibrated for driving, which captures attention demand, visual demand, temporal demand, and situational stress. DALI has since been adopted as a standard subjective measure in driving interaction studies \cite{weidner2019interact}. On the objective side, Lateral Lane Maintenance (LLM) \cite{he2014lane, mok2017tunneled} and Speed Maintenance (SM) \cite{chrysler2010driving, lewis2011speed} quantify the degree to which a secondary task disrupts primary vehicle control. Because most prior spatial interaction studies assumed SAE Level~3--4 automation, where primary driving demands are minimal, these objective measures were seldom necessary. Under Level~2 conditions, however, the driver retains full vehicle control; accordingly, the present study adopts this dual-task framework to empirically validate whether temporally decoupled spatial referencing can mitigate cognitive friction without compromising safety under Level 2 driving.
\section{Observation Study}
To understand how drivers interact with external surroundings while driving, we conducted an observational study of real-world driving. To capture diverse pointing cases, we observed both driver- and passenger-initiated interactions. The study has following three aims: 
\begin{itemize}
    \item Explore the contexts and purposes for which drivers reference external objects;
    \item Analyze the methods drivers use to point to external objects;
    \item Identify constraints of secondary interactions while driving.
\end{itemize}

\subsection{Participants}
In total, 11 participants (O1-O11) were recruited
who had at least 3 months of actual driving experience on the road (9 male and 2 female participants, average age: 32.8 years, age standard deviation: 2.2 years, average driving experience: 95.1 months, driving experience standard deviation: 25.8 months).

\subsection{Study Setup}
\begin{figure}[tb]
  \includegraphics[width=\columnwidth]{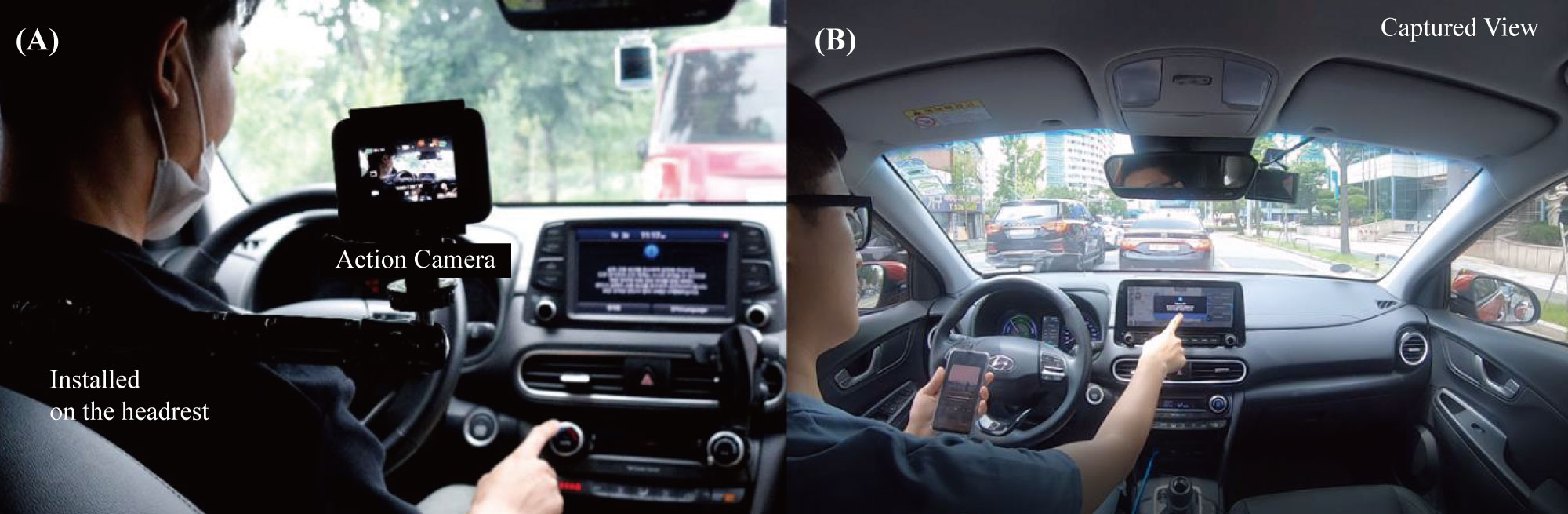}
  \caption{Study Setup: (a) Position of the camera, (b) Captured view}
  \label{fig:videoethnography}
\end{figure}

To capture their natural driving behavior and ensure they felt most comfortable during the experiment, participants drove along their usual daily routes in their personal vehicles while being recorded, with no additional restrictions. An action camera was installed on the driver's seat headrest, providing a view through the vehicle's windshield to capture the surrounding environment, including the driver's seat, passenger seat, cluster, center fascia, and dashboard (Figure \ref{fig:videoethnography}).

\subsection{Procedure}
Over a three-day period that included a weekend, participants were instructed to record at least 4 hours of video footage to capture a diverse range of driving purposes and behaviors. During an initial orientation session, participants were briefed on the use and handling of the camera. For safety and privacy, they were allowed to decide when to record, with the option to pause if they felt it interfered with driving or raised privacy concerns. After completing the video ethnography, a follow-up interview of approximately 30 minutes was conducted to clarify ambiguous actions observed in the footage and gather further insights. The entire experimental process was approved by the Institutional Review Board. 

\subsection{Data Analysis}
A total of 48.7 hours of video recordings and 5.5 hours of post-driving interview data were collected. We treated the video footage and interview transcripts as an integrated dataset, employing an iterative, mixed-methods approach grounded in thematic analysis. Initially, researchers repeatedly reviewed the video footage alongside the corresponding interview data to identify and isolate spontaneous referencing events. Through a continuous process of open coding and iterative categorization, these events were classified based on three primary dimensions: (1) contextual triggers (e.g., navigating, sharing experiences), (2) target object characteristics, and (3) specific interaction modalities employed by the occupants. By cross-referencing the observable physical pointing behaviors in the video with the drivers' subjective verbal reflections from the interviews, we iteratively refined our categories to extract overarching themes regarding their natural pointing behaviors and interaction challenges.

\subsection{Results}
\vspace{0.2cm}
\noindent\textbf{Drivers' Everyday Indications with External Objects}

\begin{figure*}[t]
  \includegraphics[width=\textwidth]{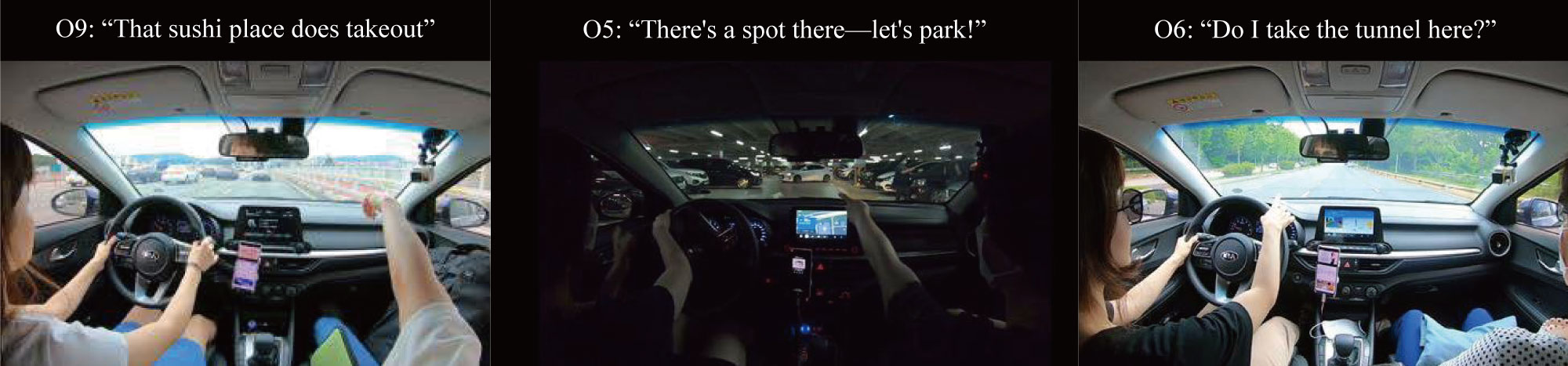}
  \caption{Participants indicating external targets}
  \label{fig:ethno_result1}
\end{figure*}

We found that the majority of drivers (8 of 11, 72.7\%) indicate surrounding objects or locations during their daily driving routines  (O1-4, O6, O8-10) (Figure \ref{fig:ethno_result1}). The primary types of referenced objects were categorized into the following five groups: 
\begin{itemize}
    \item Specific points on the road, such as intersections, turn-off points, or alley entrances;
    \item Informational media, including road signs and billboards
    \item Fixed objects, such as buildings, stores, structures, or roadside trees
    \item Dynamic objects, including vehicles, pedestrians, or animals
    \item Non-specific areas, indicated by phrases like "over there," "this block," or "around that area."
\end{itemize}

The indications were primarily observed in three contexts: route navigation (e.g., O6: "Can we make a U-turn there?"), curiosity (e.g., O3: "What kind of car is that?"), and experience sharing (e.g., O9: "That restaurant is really good."). Such use of external information was particularly frequent when drivers were accompanied by passengers.

\vspace{0.2cm}
\noindent\textbf{Prioritizing Timely and Contextual Indication Over Accuracy}

Since external objects quickly pass out of view, drivers consistently prioritized timely and contextual indications over precise verbal 
descriptions. Rather than providing detailed descriptions of external objects, drivers frequently relied on deictic expressions such as 
``that'' or ``there,'' accompanied by gestures---including nodding, finger pointing, or hand movements---to convey approximate direction (``left'' or ``right'') and distance (``near'' or ``far'') relative to the road. Drivers sometimes employed subtle pointing gestures---such as briefly lifting only the index finger. These behaviors suggest that, in practice, a rough spatial reference was sufficient for communication between occupants, making precise target specification largely unnecessary in dynamic driving contexts.

Among the observed gestures, finger pointing was the most frequently used method. During the post-study interviews, all drivers who 
exhibited pointing behaviors (O1--4, O6, O8--10) confirmed that finger pointing felt like the most intuitive and straightforward 
modality for spatial communication, corroborating the findings of Tscharn et al. \cite{tscharn2017stop}. 

\vspace{0.2cm}
\noindent\textbf{Considerations of Pointing in Real-Driving Situations}

While driving, drivers frequently operate navigation systems, select music, answer calls, or converse with passengers. However, for tasks requiring prolonged attention, such as text input into infotainment systems or processing large amounts of information, our observations showed that drivers actively defer these demands to safe moments, such as waiting at traffic signals. For example, during a destination reset process, one driver encountered a long list of similar location names, which led them to pull over to the roadside entirely to safely resolve the issue.

Because driving safety and continuous vehicle control remain the absolute priority, drivers show a reluctance to release the steering wheel for secondary tasks. Even when indicating external objects, they preferred subtle movements—such as briefly lifting an index finger—and sometimes omitted physical gestures altogether if the conversational context was sufficiently obvious. This highlights the need for simple, quick gestures that minimize distractions while maintaining a firm grip on the steering wheel.

POIs on the road pass by quickly, with nearby objects leaving the driver's field of view even faster. As drivers need to keep their eyes on the road, it becomes challenging to turn their heads and follow passing objects. Drivers and passengers often referred to missed objects using temporal expressions like "that one earlier" or "just now" to describe them. Also, they struggled to communicate when POIs were obscured by obstacles, depending on their seat position. Given these, it is also beneficial to provide access to information from previous moments.

\section{Design of Point and Select}

\begin{figure}[b]
  \includegraphics[width=\columnwidth]{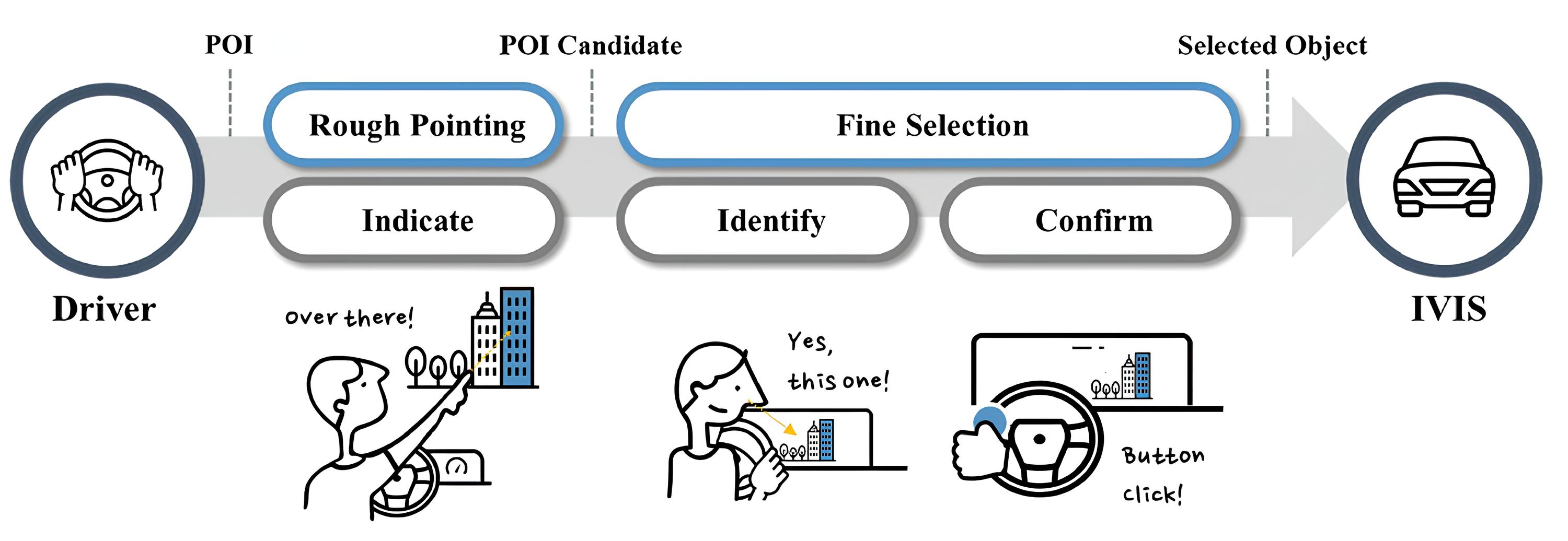}
  \caption{Main concept of Point and Select}
  \label{fig:main concept}
\end{figure}

To support the integration of external surrounding objects into the IVIS while driving, we designed a spatial interaction method named Point and Select. The design concept of Point and Select is illustrated in Figure \ref{fig:main concept}. The Point and Select interaction consists of two steps: a rough pointing phase and a fine selection phase. In the rough pointing phase, drivers indicate an external object using their index finger—a natural human gesture—allowing them to rapidly and intuitively select the intended object. Based on this input, our prototype recommends candidates at the point the driver indicated. To address the challenge of identifying specific objects in a complex and fast-moving environment, we introduced a second step: fine selection. The fine selection phase allows the driver to clearly identify the initial POI candidates, make adjustments if the selection is unintended or needs to be changed, and finally confirm the choice. The following sections detail our design considerations and features for Point and Select, derived from observation studies and literature review.

\subsection{Features and Design Considerations for Point and Select Interaction}

\vspace{0.2cm}
\noindent\textbf{Pointing Gesture: Intuitive Behavior for Low-Cognitive Load}  

In driving situations, as cognitive load can negatively affect safety and driving performance, interactions should be designed to be intuitive. Pointing gestures, rooted in natural human behavior, were identified through observation studies as an intuitive input method. These gestures are straightforward and unlikely to confuse users, making them suitable for dynamic driving contexts.

\vspace{0.2cm}
\noindent\textbf{Rapid and Non-Precise Input: Approximate Indication via Rough Pointing}  

As directly observed in our video ethnography, external targets pass out of view rapidly, prompting drivers to prioritize timely, approximate indications over precise spatial targeting. In fact, mid-air direct pointing is challenging for drivers to perform accurately while driving \cite{graichen2019evaluation, gomaa2020studying}, as requiring precise selection can distract from the primary driving task and increase response time. To address this, we adopted a rough selection approach to minimize distraction and enable quick input at the desired moment. The system identifies the most likely object by evaluating the proximity of the pointing direction to an object and its distance from the driver, even if the extended line of the finger does not pass exactly through the object.

\vspace{0.2cm}
\noindent\textbf{Clear Visual Cues: Cursor and Perspective Alignment}

\begin{figure}[t]
\centering
  \includegraphics[width=0.8\columnwidth]{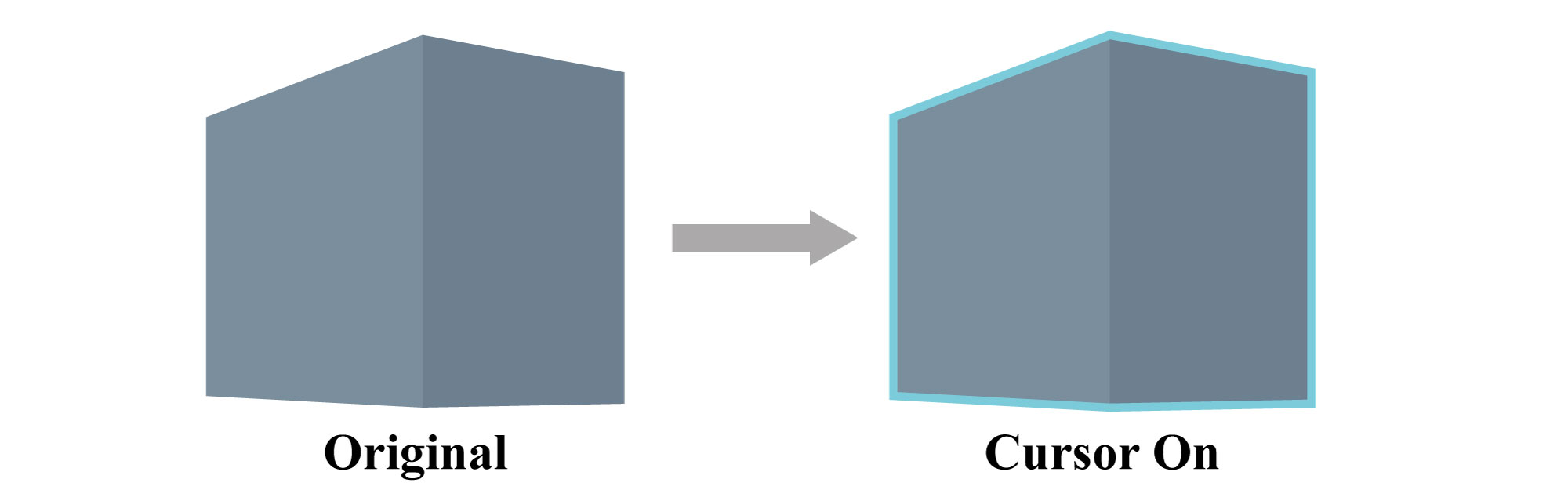}
  \caption{Visual cursor with highlighted outline on digital instrument cluster}
  \label{fig:cursor}
\end{figure}

Since mid-air gestures like finger-pointing lack tactile feedback, visual feedback is essential \cite{monnai2014haptomime,saponas2009enabling,martinez2017}. To ensure drivers can quickly and clearly perceive the selected POI, the system highlights the object’s outline, referred to as the Cursor (Figure \ref{fig:cursor}), allowing identification at a glance even from a distance. We avoid putting indicators like arrows due to potential confusion caused by overlapping objects.

\begin{figure}[b!]
  \includegraphics[width=\columnwidth]{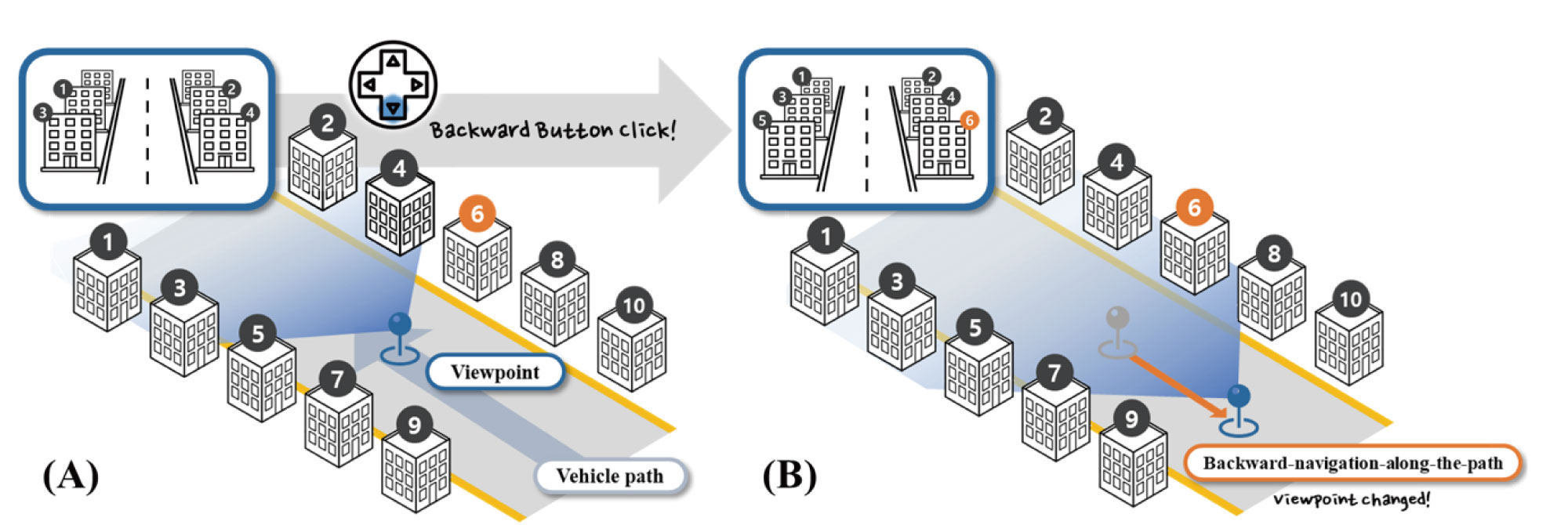}
  \caption{Concept of ‘Route-based backward navigation’: (a) When the driver passes by POI 6, (b) they can navigate backward along the
path to select POI}
  \label{fig:backward}
\end{figure}

After pointing, two options for the initial driver-facing view were considered: the cursor-centered view and the driving-direction view. The cursor-centered view places the selected object at the center of the screen but compromises spatial context, particularly for nearby objects. In contrast, the driving-direction view aligns with the vehicle's trajectory, offering a forward-facing perspective that more effectively conveys the selected object’s position within the environment. The driving-direction view was chosen as the default for its clarity and natural alignment with the driver’s perspective (Scenes inside the rounded rectangles in Figure \ref{fig:backward}). Furthermore, to maintain spatial awareness during viewpoint transitions when navigating between objects, an animation effect was added to visually guide the cursor's movement along the route \cite{therrien2010spatial}.

\vspace{0.2cm}
\noindent\textbf{Temporal Decoupling and Fine Selection: Route-Based Backward Navigation}

In dynamic driving scenarios, objects quickly pass by, causing drivers to miss or misselect targets during the initial pointing phase. To address this spatiotemporal mismatch, the interface needs to allow retrospective navigation without increasing cognitive load. Furthermore, as our observations highlighted drivers' strong reluctance to release the steering wheel, the interaction must minimize physical disruption. To achieve this, instead of free-cursor exploration, we first implemented a snapping-based selection approach, where the cursor automatically jumps between valid candidate objects. Next, to manipulate the spatiotemporal navigation, we evaluated two browsing methods: driver-centric rotary navigation and route-based backward navigation. \textit{Driver-centric rotary navigation} anchors the view to the vehicle’s position and rotates horizontally, which can cause disorientation and make distant objects harder to target. \textit{Route-based backward navigation} moves the cursor backwards along the actual driving path, intuitively matching the driver's recent spatial memory. Therefore, the route-based backward navigation method was selected (Figure \ref{fig:backward}). Crucially, to ensure continuous vehicle control, this spatiotemporal navigation and the final confirmation are operated entirely via the steering wheel buttons. This design enables drivers to comfortably browse and select past targets at a self-paced moment without breaking their grip on the wheel.

\section{Implementation}
In this section, we describe the implementation of driving simulation environment and the Point and Select interaction technique.

\begin{figure*}[t]
  \includegraphics[width=0.9\textwidth]{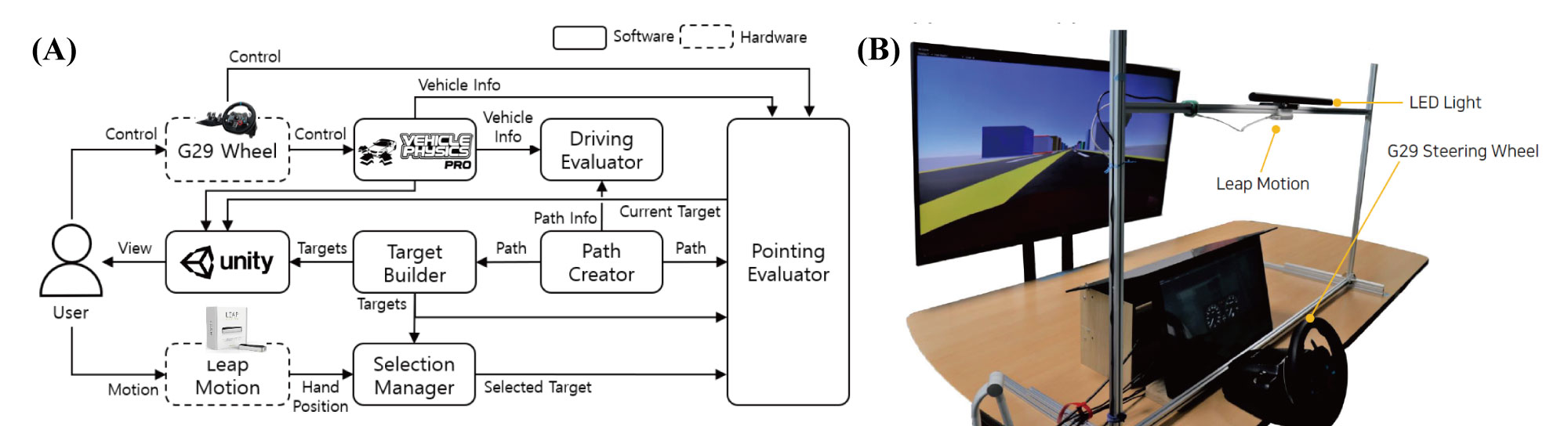}
  \centering
  \caption{Prototype of Point and Select: (A) Software Architecture, (B) Hardware Configuration}
  \label{fig:prototype system}
\end{figure*}

\subsection{Simulation Environment}
3D simulated driving environment was developed using Unity 2021 LTS (Figure \ref{fig:prototype system} (A)). To create realistic vehicle dynamics, Vehicle Physics Pro, an automotive physics engine, was primarily utilized, allowing drivers to experience car movement similar to real-world scenarios. To enhance immersion, steering sensitivity, resistance, and road surface vibration were fine-tuned based on the Logitech G29 steering wheel.

\begin{figure}[t]
  \includegraphics[width=\columnwidth]{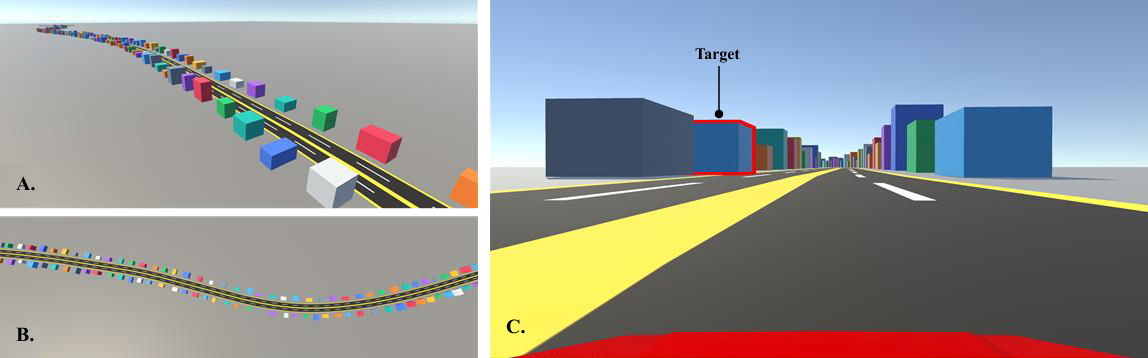}
  \caption{Configuration of the Driving Route: (A) Perspective view, (B) Top view, (C) Driver's view (the red outline highlights the target to be selected).}  
  \label{fig:driving course}
\end{figure}

The road layout was designed using the Bézier Path Creator, combining irregularly arranged straight sections with smoothly curved ones, while buildings were randomly positioned on both sides beginning 100 meters from the starting point of the course. The road was designed to be 3.3 meters wide per lane with two lanes, following the Korean city road regulations. This design allowed drivers to quickly adapt to the lane width and their sense of speed. 
The virtual vehicle was modeled with a width of 1.86 meters, representing the dimensions of a mid-size sedan. Considering that the spacing and size of buildings influence POI density and, consequently, users' task performance, we designed a simulated city environment. The buildings were randomly generated with dimensions ranging from 10 × 10 × 8 m³ to 20 × 20 × 15 m³, positioned 5–8 meters from the road to reflect typical urban layouts. Furthermore, in adherence to the regulations on the minimum spacing between buildings, the distances between structures ranged from 10 to 30 meters, while maintaining the building density in urban areas. Both the width of the road and the dimensions of the building were scaled proportionally to the vehicle. The layout of the simulated driving course is shown in Figure \ref{fig:driving course}. Real-time logging was implemented to record data such as vehicle position, speed, time, and target selection activities throughout the simulation. 

A 55-inch display was used to showcase the driving environment beyond the vehicle’s windshield, designed to resemble the typical view seen through a front windshield during urban driving. The building density within the field of view was also adjusted to align with real-world urban road environment. The upper half of a 22-inch display served as a Driver Information Center (DIC), simulating a traditional instrument panel while also displaying selected POIs, which will be discussed. The positions of the 55-inch display and the steering wheel were installed based on the scene generated by the driving simulation system. Considering the characteristics of Korean vehicles, where the driver’s seat is located on the left, the steering wheel and the DIC display were positioned slightly to the left of the center.

A Leap Motion controller was installed above the driver using a custom aluminum profile to ensure effective hand recognition. To optimize recognition accuracy, bright LED lighting was installed above the Leap Motion sensor to maintain consistent illumination in the environment.

Figure \ref{fig:prototype system} illustrates the software architecture and hardware configurations of the simulated driving environment and interface.

\subsection{Point and Select Interface}
To implement Point and Select interaction scenario, we broke it down into three steps: Indicate, Identify, and Confirm.

\vspace{0.2cm}
\noindent\textbf{Rough Pointing Phase: Indicating the Approximate Location of the Target}

\begin{figure}[t]
  \includegraphics[width=0.9\columnwidth]{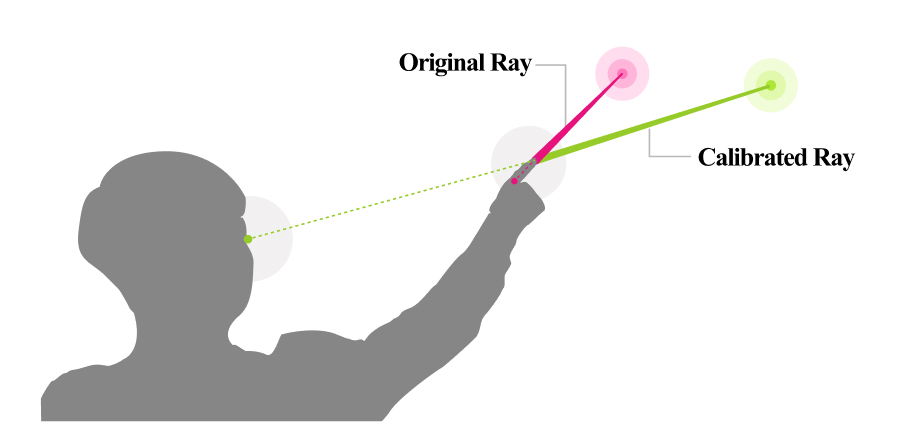}
  \centering
  \caption{Original ray and calibrated ray}
  \label{fig:hand model}
\end{figure}

The rough pointing phase allows drivers to quickly and intuitively provide initial input based on approximate distance and direction. To identify the POI, the Leap Motion controller was used to track the driver’s hand movements and recognize the finger direction. Accuracy was further enhanced by activating robust tracking mode and illuminating the user's hand with a white LED lamp, creating a bright lighting environment. The driver’s finger-pointing gesture was detected after pressing the activation button. The finger-pointing gesture algorithm was developed based on the index finger of the user's right hand, with the original ray constructed by connecting the third joint to the fingertip (Figure \ref{fig:hand model})

\vspace{0.2cm}
\noindent\textit{Calibration}
 
Calibration is necessary because gaze alignment and binocular disparity vary between individuals, and the starting points of the gaze and the right index finger do not always match. The ray drawn from the pointing gesture was refined through calibration. The calibration process began when the simulated driving environment was first activated, allowing the user to enter calibration mode by pressing a designated button. In this mode, five evenly spaced red dots appeared across the screen from the left to the right edge. While maintaining forward focus, drivers pointed at each dot with their index finger and pressed the button when they believed they had accurately pointed at the dot. This process was repeated five times for each dot, totaling 25 trials, which typically completed within two minutes. Note that this calibration needs to be performed only once per driver, as the resulting profile can be saved and reloaded for future sessions. These data were used to calculate where the finger was actually pointing on the simulator screen.

A second-step correction is to snap the closest object, even if the calibrated ray does not directly intersect it. This was achieved by calculating how closely the calibrated ray pointed toward an object. An array of vectors was generated from the ray's origin (the index finger's 0th joint) to each object on the road. The angles to the calibrated ray were calculated using the dot product of normalized vectors, with values near 1 indicating better alignment. The object with the smallest angle and shortest distance within a defined range was identified as the driver’s target, as shown in Figure \ref{fig:hand model}.

\vspace{0.2cm}
\noindent\textbf{Fine Selection Phase: Identifying the Candidates and Confirming the POI}

In the fine selection phase, displaying the POI candidate selected in the rough pointing phase is crucial, as the target disappears and mid-air finger gestures lack feedback \cite{cornelio2017agency,monnai2014haptomime,saponas2009enabling}, often leaving the driver uncertain about their pointing. To address this, the display needs to ensure rapid feedback and a clear view while driving, without requiring the driver to divert attention from the road \cite{sauras2017voge}. Among in-vehicle displays, the center fascia offers a large interface but requires the driver to look away from the road. Head-Up Displays allow forward focus but are limited in covering wide areas for POIs, may obstruct forward vision when displaying the full view, and lack clarity due to their transparency. The DIC, increasingly replaced by digital displays, combines the familiarity of presenting driving information with the convenience of a large display like the center fascia, while allowing drivers to maintain forward view \cite{rao2014design}. Therefore, we used DIC to display the front view and POI candidates, with the first step showing an image of the scene at the time of rough pointing, enabling drivers to quickly identify the scene without spatial cognitive dissonance. The first POI candidate selected through rough pointing was marked with a visual cursor to ensure quick and clear recognition (Figure \ref{fig:cursor}).

\begin{figure}[t]
  \includegraphics[width=\columnwidth]{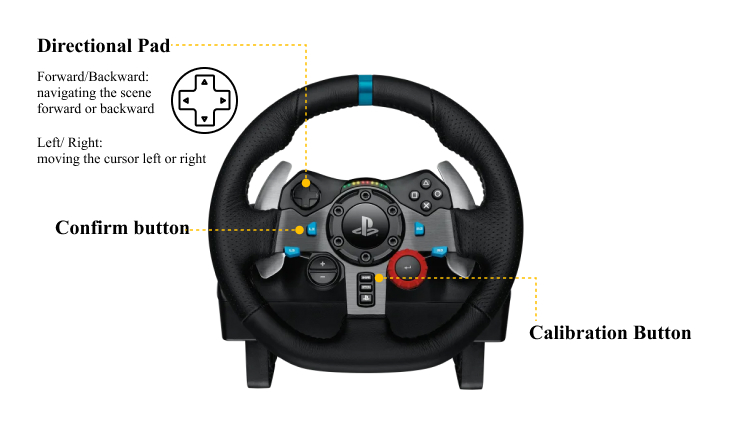}
  \caption{Button interfaces on the steering wheel}
  \label{fig:prototype Button}
\end{figure}

The final step is to allow drivers to adjust the cursor if needed and confirm the selection to input the object. Given the unpredictable nature of dynamic driving environments, we prioritized ensuring that the driver keeps at least one hand on the steering wheel to respond quickly and can return their hand to the steering wheel immediately after performing secondary tasks. We used arrow key buttons on the steering wheel to manipulate the on-screen cursor, providing an input method that allows for convenient eye-free operation while driving. The down-arrow key enabled backward navigation along the path, allowing drivers to select previously encountered objects, while the up-arrow key intuitively moved the cursor forward along the path (Figure \ref{fig:backward}). Similarly, the left- and right-arrow keys allowed the cursor to move across the road. These keys enabled users to position the cursor on the target object, after which the driver could finalize the selection by pressing a designated confirmation button. To enhance spatial perception during this navigation, animation was incorporated to visually represent the viewpoint's movement along the path \cite{therrien2010spatial}. The described functionality was implemented using the Logitech G29 steering wheel, with specific functions assigned to each button for these operations (Figure \ref{fig:prototype Button}).

\begin{figure*}[t]
  \includegraphics[width=0.95\textwidth]{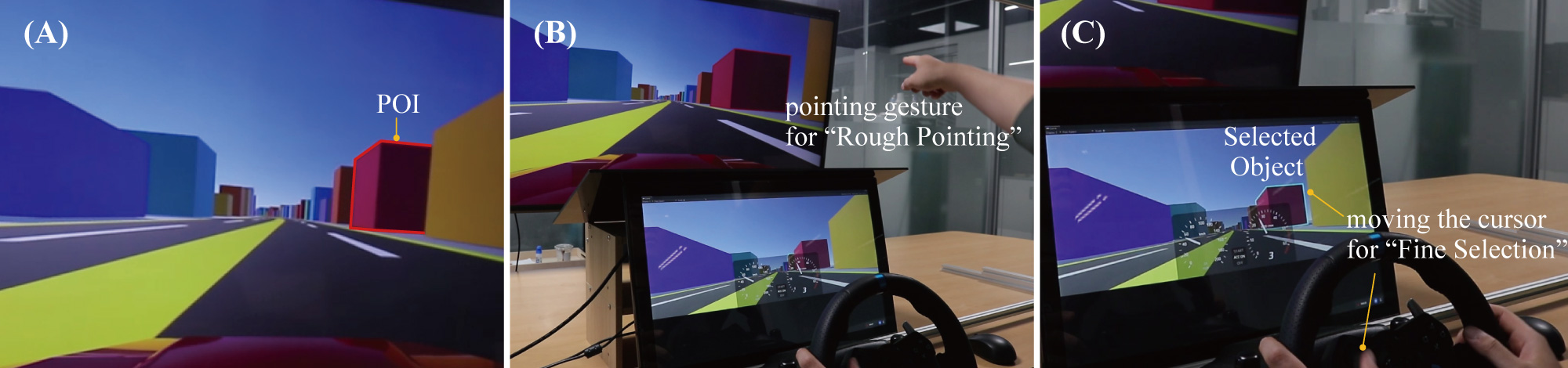}
  \centering
  \caption{Implementation of Point and Select Interaction Scenario: (A) target assign, (B) Rough Pointing, (C) Fine Selection}
  \label{fig:implemented scenario}
\end{figure*}

\vspace{0.2cm}

The implemented Point and Select interaction in the simulated environment is illustrated in Figure \ref{fig:implemented scenario}. When a driver identifies their POI on the front display (Figure \ref{fig:implemented scenario} (A)), they can quickly select the target direction by roughly pointing with their finger (Figure \ref{fig:implemented scenario} (B)). The DIC then displays the POI candidates, allowing the driver to fine-tune their selection using a button on the steering wheel (Figure \ref{fig:implemented scenario} (C)).

\section{User Study}
The primary objective of user study is to evaluate the performance and usability of Point and Select interaction, at different urban driving speeds and its impact on the primary task. Through this evaluation, we aim to determine whether Point and Select interaction is acceptable while driving.

\subsection{Participants}
We recruited participants with over three months of actual driving experience in the urban road environment. This was to ensure that they were fully aware of the local road regulations and capable of explaining the experience of using Point and Select in comparison with their typical driving experience.. In addition, only those who could distinguish colors normally (i.e., no color blindness) were recruited because the experiment required color discrimination. The final set of participants comprised 12 (six male and six female, P1 - P12) drivers. Their average age was 34.6 years (S.D = 1.7 years), and their average driving experience was 70.7 months (S.D = 9.5 months). We also received each participant's consent for recording the process and using it after anonymization. 

\subsection{Study Setup}
A user study was conducted in a simulated driving environment considering safety and difficulty of testing scenarios in real-world road environments. Before each participant took part in the experiment, the pointing gesture was calibrated as described in the previous section. Additionally, the positioning of the chair, pedals, screens, and steering wheel, along with its resistance, was adjusted for each participant to closely replicate the feel of an actual driving experience. Cameras for video and audio recording were setup at a distance to avoid interfering with the driver's vision or behavior.

\subsection{Task}
To evaluate the applicability of the Point and Select interaction while driving in an urban setting, we defined two conditions: Condition 1 (C1), the baseline condition where drivers performed only the primary task of driving along a path, and Condition 2 (C2), the experimental condition where drivers performed both the primary task and the Point and Select interaction. Since driving speed is a critical factor influencing performance, we introduced three speed levels: 30 km/h, representing the speed limit for residential roads; 50 km/h, the speed limit for general urban roads; and 70 km/h, the speed limit for vehicle-only roads. As a result, the two experimental conditions, C1 and C2, were divided into three driving speed levels.

To be specific, in C1, participants were instructed to drive on the course for 5 minutes, maintaining their position in the center of the first lane and adhering to a specified speed range. In C2, participants performed an additional task of inputting target objects while driving under the same conditions as in C1. The target objects were randomly assigned within the driver's visible range, selected from buildings on both sides of the road, and these buildings were highlighted with a red outline displayed on the driver's front screen (Figure \ref{fig:driving course} (C)). The drivers were asked to select the target with Point and Select. The outcome was denoted as success if fine selection was successful, wrong if fine selection failed, and missed if fine selection was not performed within the time limit or rough pointing was missed. A new target was assigned every 3–6 seconds when a target was either successfully selected or missed. In total, the system generated at least 25 targets.

\subsection{Procedure}
Participants first completed a questionnaire about their driving experience, including their routine engagement in secondary tasks and usage of IVIS. Afterward, participants were introduced to the Point and Select interaction and the simulated driving environment through a 10-minute orientation.

Participants drove freely for about 10 minutes to familiarize themselves with the steering wheel's feedback and the sense of speed in the simulation. This was followed by another 10-minute session practicing the Point and Select interaction while driving, allowing them to get accustomed to selecting external objects. Participants were instructed to stop the experiment immediately if they experienced dizziness. After adapting to the Point and Select interaction and simulated environment, participants evaluated the environment. Then, for each of the six conditions, they completed a five-minute task and a three-minute survey, followed by a two-minute break. They answered the DALI questionnaire in Condition 1 (C1) and both the DALI and NASA-TLX questionnaires in C2. The order of speed levels was randomized for each participant.

Finally, a 30-minute semi-structured interview was conducted, focusing on driving performance and cognitive load, observations of unusual behaviors, comprehensive assessments of the Point and Select interaction, comparisons with routine secondary tasks, and suggestions for improvement. The entire session, lasting approximately 120 minutes per participant, was recorded on video.

\subsection{Measurements}

We assessed primary driving degradation and secondary interaction usability using the following measures.

\vspace{0.2cm}

\noindent\textbf{Evaluation of simulated driving environment} 

Prior to the main experiment, the Simulator Sickness Questionnaire (SSQ) \cite{kennedy1993simulator} was administered to assess the severity of nausea, oculomotor disturbances, and disorientation to ensure the 3D environment did not induce cyber-sickness.

\vspace{0.2cm}

\noindent\textbf{Driver's performance} 

\noindent\textit{Primary Driving Task Performance:} We calculated LLM as the standard deviation of the distance from the vehicle's centre to the lane centre, and SM as the root-mean-square deviation from the target speed based on logged vehicle data. 

\noindent\textit{Primary Driving Task Load:} The DALI questionnaire \cite{pauzie2008method} was administered on a 7-point Likert scale after each of the six conditions to assess the cognitive workload of driving.

\vspace{0.2cm}

\noindent\textbf{Interaction usability} 

\noindent\textit{Secondary Task Performance:} We measured Success Rate (SR), defined as the ratio of successful selections to total attempts, and Task Completion Time (TCT), the interval from the target's appearance to successful confirmation. 

\noindent\textit{Secondary Task Load:} The NASA-TLX questionnaire \cite{hart1988development} was administered on a 7-point Likert scale after the C2 conditions to assess the cognitive workload imposed specifically by the Point and Select interaction. 

\noindent\textit{Secondary Task Comparison:} During the post-interview, participants ranked their typical real-world secondary tasks in order of cognitive difficulty relative to the Point and Select interaction.

\vspace{0.2cm}
\noindent\textbf{Statistical Analysis} 

Statistical analyses were conducted using non-parametric tests, as the data—particularly subjective workload metrics (DALI, NASA-TLX) and the relatively small sample size ($N=12$). To evaluate the differences between the two conditions (C1 vs. C2) at each speed level, we employed the Wilcoxon signed-rank test. For comparisons across the three speed levels (30, 50, and 70 km/h) within the same condition, we utilized the Friedman test. Where significant main effects were found, Dunn's test was applied for post-hoc pairwise comparisons.
\section{Result}

\begin{table}[t]
\centering
\begin{tabular}{|c|c|c|c|c|}
\hline
 & \textbf{Nausea} & \textbf{Oculomotor} & \textbf{Disorientation} & \textbf{Total Score} \\ \hline
\textbf{Average} & 15.11 & 22.74 & 26.68 & \textbf{6.50} \\ \hline
\textbf{S.D} & 12.60 & 18.57 & 21.61 & 4.80 \\ \hline
\end{tabular}
\caption{Results of SSQ questionnaire}
\label{tab:Table2}
\end{table}

\subsection{Validation of the Simulated Driving Environment}

The result of SSQ is summarized in Table \ref{tab:Table2}. After getting used to the simulated driving environment, all participants reported no issues with using the simulated driving environment, and the majority selected "no symptoms (0)" for almost all items. As a result, the SSQ total score averaged 6.3 (SD: 4.80), which is significantly lower than the threshold score of 32 \cite{kennedy1993simulator}. Thus, the effect of cyber-sickness was not severe enough to influence the experimental results. Participants noted that the steering wheel's force feedback and the accompanying sounds closely resembled the actual driving experience, enhancing their sense of immersion. P2 mentioned "At first, the driving environment felt a bit different from my car because of the size, but I got used to it quickly. The weight and feel of the steering wheel were pretty similar to my car’s," while P4 noted "I didn’t even need to look at the speedometer because I could guess the speed just by the exhaust sound."

\begin{figure}[b]
  \centering
  \includegraphics[width=0.95\columnwidth]{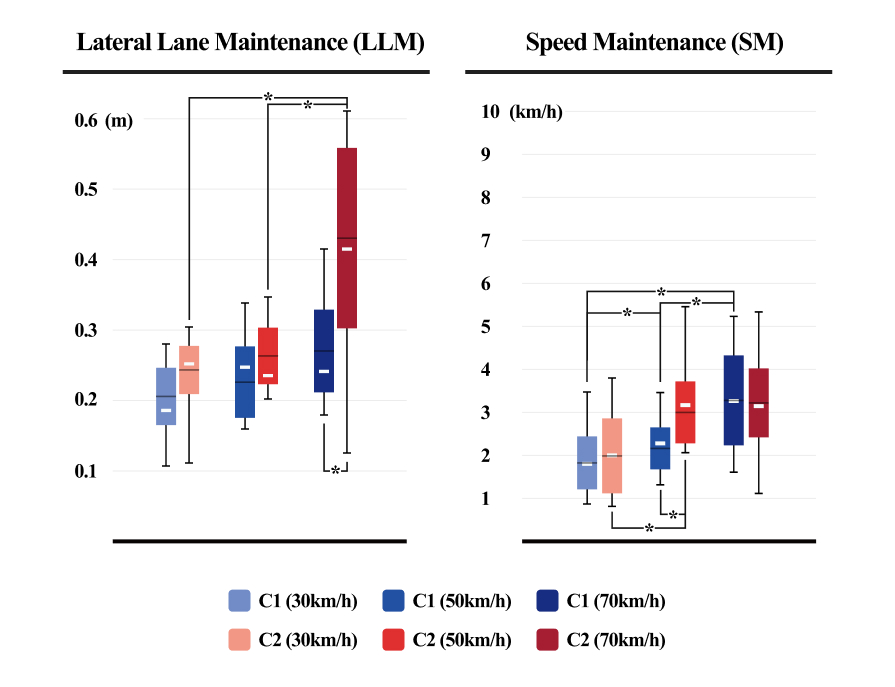}
  \caption{Results of Lateral Lane Maintenance (Left), Results of Speed Maintenance (Right)}
  \label{fig:LLM}
\end{figure}

\begin{figure*}[t]
  \includegraphics[width=0.8\textwidth]{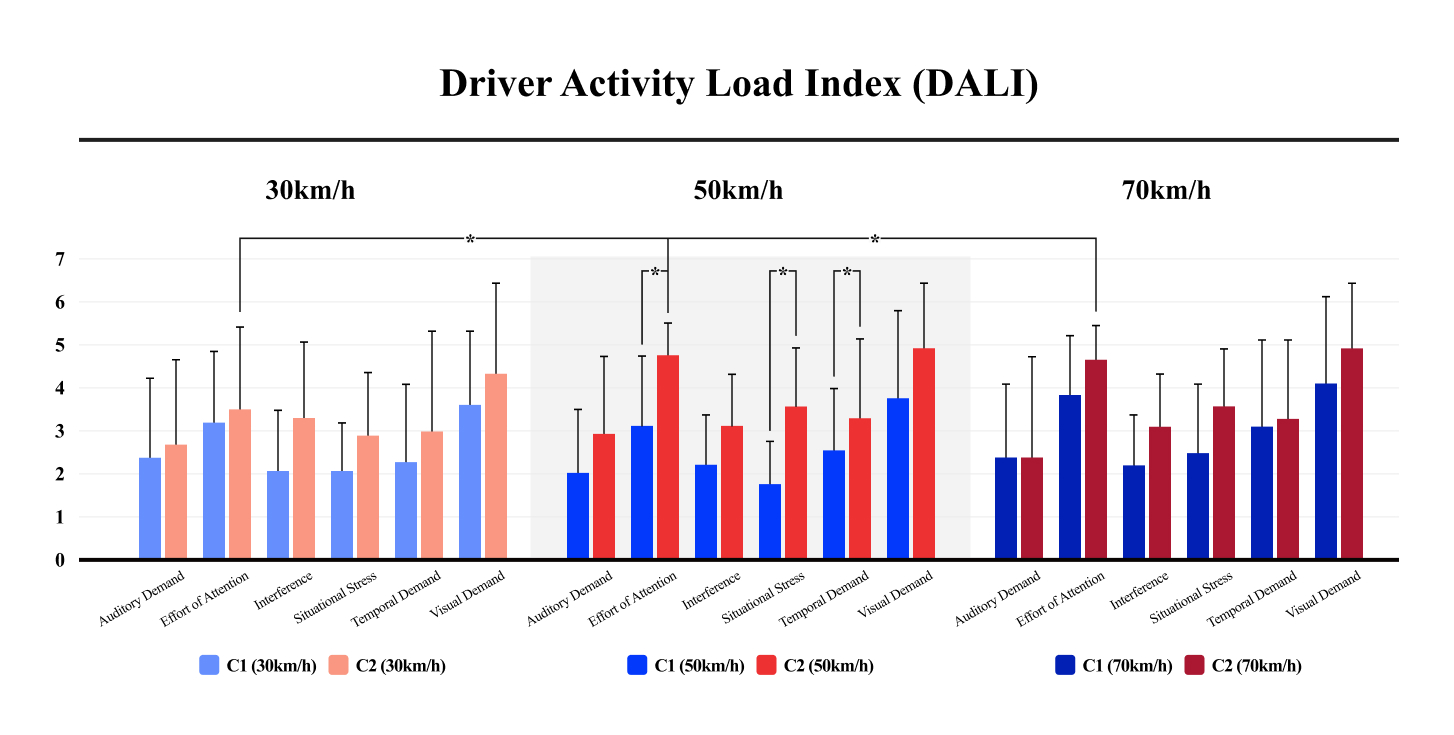}
  \centering
  \caption{Results of DALI questionnaires in 3 different speed levels (30 km/h, 50 km/h, 70 km/h)}
  \label{fig:DALI}
\end{figure*}

\subsection{Driver Performance}
\vspace{0.2cm}
\noindent\textbf{Primary Driving Task Performance}

\noindent\textbf{Primary task performance (LLM).}
Figure~\ref{fig:LLM} summarizes lane-keeping (LLM) and speed maintenance (SM).
In both conditions (C1: without Point-and-Select and C2: with Point-and-Select), LLM increased with speed, but \emph{only} at 70~km/h did C2 show significantly worse lane-keeping (higher LLM) than C1 ($p=.02$); 30 and 50~km/h did not show significant difference between conditions. Within C2, 70~km/h $>$ 30/50~km/h (each $p<.05$).

\medskip
\noindent\textbf{Speed maintenance (SM).}
SM also increased with speed. A between-condition difference appeared \emph{only} at 50~km/h (C2 $>$ C1, $p=.02$); there was no condition effect at 30~km/h ($p=1.00$) or 70~km/h ($p=.06$). Within conditions, SM increased with speed in C1 (30 $<$ 50 $<$ 70; all $p<.05$), whereas in C2 only 30~km/h differed from 50 and 70~km/h (each $p=.01$), with 50~km/h $\approx$ 70~km/h ($p=.99$).

Post-session interviews revealed that this was influenced by a gear shift that occurred around 50 km/h, which had a greater impact in C2 compared to C1:
{\itshape
``At 50~km/h, the pedal response felt different due to a shift-up. When I was just driving, I focused on that; with Point-and-Select, I paid \textbf{less} attention to speed.'' }(P12)

{\itshape``Around 50~km/h the car would pause and then pick up as the gears changed. In that condition I kept checking the speedometer because of the speed limit.''} (P7)

Given that LLM is driven by steering control and visual attention, whereas SM depends on pedal operation and speed familiarity, Point-and-Select shows its impact in more demanding situations---those requiring precise steering (high speed) or that disrupt the driver’s accustomed sense of speed.

\vspace{0.2cm}
\noindent\textbf{Primary Driving Task Load}

The primary driving task load assessed with DALI is summarized in Figure~\ref{fig:DALI}.
Despite the additional load introduced by Point-and-Select (C2), a between-condition difference emerged \emph{only} at 50~km/h.
Participants reported that once adapted, the interaction could be executed with brief glances (\emph{“The interaction itself did not interfere much with driving; once adapted, I could use it quickly with a glance.”}—P10).
At 50~km/h, C2 exceeded C1 in \emph{effort of attention} ($p=.02$), \emph{situational stress} ($p<.001$), and \emph{temporal demand} ($p=.04$).
This pattern is consistent with the previously noted shift-up region near 50~km/h, and with post-interview accounts emphasizing sudden target appearance during speed maintenance (P4, P5, P7, P8, P11, P12).
As one participant noted, \emph{“Selections were not particularly difficult at the other speeds, but at 50~km/h I felt momentarily flustered when a target appeared while cornering.”} (P5)

Across speeds, Point-and-Select showed greater impact on \emph{effort of attention} and \emph{situational stress} as speed increased.
In C2, \emph{effort of attention} differed by speed (overall test: $p=.05$; post hoc: 30--50~km/h $p=.02$, 30--70~km/h $p=.04$).
For \emph{situational stress}, the overall test was significant ($p=.03$), but Dunn’s post-hoc comparisons did not yield significant pairwise differences.



\begin{figure}[b]
  \centering
  \includegraphics[width=0.95\columnwidth]{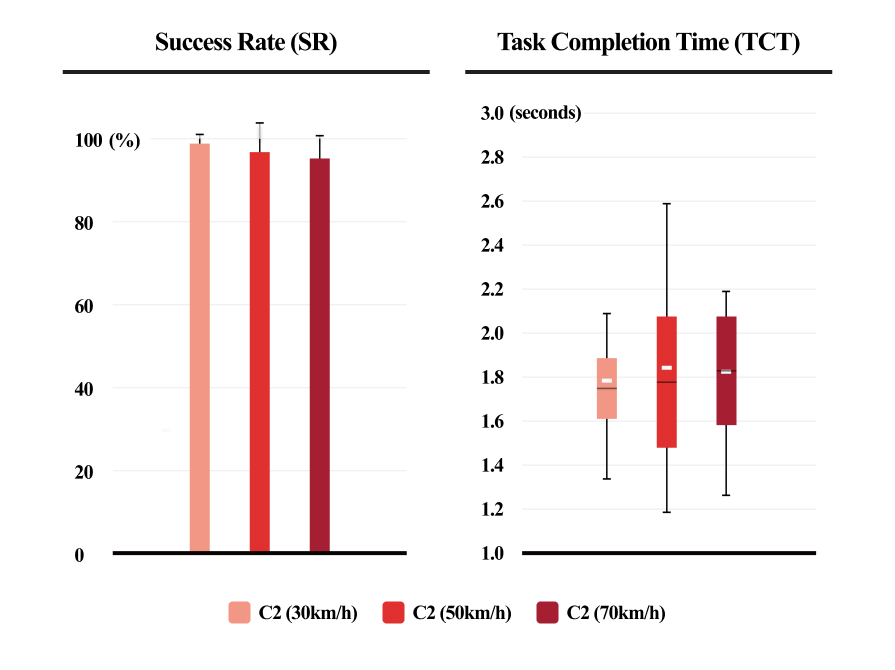}
  \caption{Results of Success Rate (SR) and Task Completion Time (TCT) in C2 (from left to right: 30 km/h, 50 km/h, 70 km/h)}
  \label{fig:SR}
\end{figure}

\subsection{Interaction Usability}

\vspace{0.2cm}
\noindent\textbf{Secondary Task Performance}

The average TCT and the average SR for three speed levels were summarized as shown in Figure \ref{fig:SR}. The average TCT was measured at 1.79 seconds (SD: 0.25 seconds) at 30 km/h, 1.84 seconds (SD: 0.39 seconds) at 50 km/h, and 1.82 seconds (SD: 0.40 seconds) at 70 km/h. The average SR was measured at 98.7\% (SD: 2.38\%) at 30 km/h, 96.7\% (SD: 6.73\%) at 50 km/h, and 95.17\% (SD: 5.35\%) at 70 km/h. The overall average for all three speed levels of TCT and SR were 1.82 seconds (SD: 0.36 seconds), and 96.9\% (SD: 5.35\%), respectively. Both the TCT and the SR according to speed were not statistically different. P11 mentioned, “I don’t think the difficulty of selection depending on the speed itself changes." Drivers quickly made rough selections at the intended timing and performed precise selections when they had more capacity or when the primary task allowed as P9 mentioned: “It wasn’t a big burden after I got used to it. Once I pointed at it (rough selection) and the marker appeared on the screen, I could complete the selection after turning the curve.” 

\begin{figure}[t]
  \centering
  \includegraphics[width=\columnwidth]{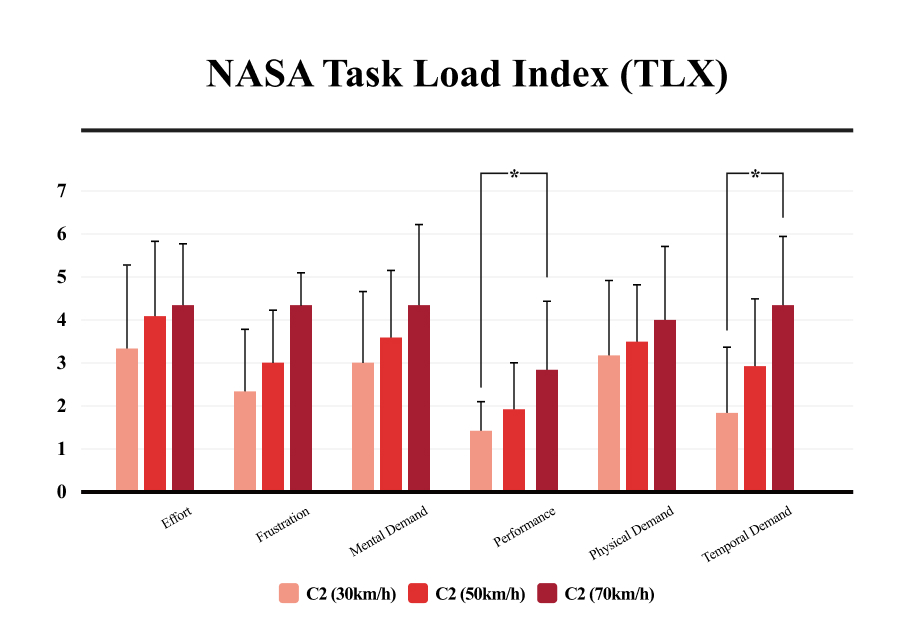}
  \caption{Results of NASA TLX questionnaires in C2 (from left to right: 30 km/h, 50 km/h, 70 km/h)}
  \label{fig:TLX}
\end{figure}

\vspace{0.2cm}
\noindent\textbf{Secondary Task Load}

Secondary task load evaluated by NASA TLX is summarized in Figure \ref{fig:TLX}. Among six metrics, there were significantly differences in temporal demand ($p$ = 0.01 < 0.05, post-hoc: $p_{\ 30-70\ km/h}$ = 0.03 < 0.05) and performance ($p$ = 0.01 < 0.05, post-hoc: $p_{30-70\ km/h}$ = 0.03 < 0.05). Increase of temporal demand was affected by the pressure to return to driving more quickly as the speed increased. Regarding this, P10 noted, “It’s not so hard to select them all. But I think I was in a hurry when the speed was fast because it’s hard to look away from the front for a long time.” Performance became negative (as a lower score indicates better performance) as the speed increased, while the SR indeed showed no significant difference (Figure \ref{fig:SR}). Drivers felt they could have performed better. Overall, participants expressed strong confidence that they could use the Point and Select interaction effectively with more practice in the simulated driving environment. P2 noted, "I think I’ll be able to get it all right after just a few more times," while P4 added, "If it were my car, I would have done better."

\vspace{0.2cm}
\noindent\textbf{Comparing with Conventional Secondary Task Load}

Participants' common secondary tasks included checking navigation maps, entering GPS destinations, making calls, selecting songs, skipping tracks, texting, and using voice commands. All participants (P1-P12) ranked tasks in order of cognitive difficulty, from easiest to hardest: clicking shortcut buttons, using shortcut voice recognition, selecting external objects via the Point and Select interface, finding information on maps or lists (even after voice recognition), and entering text. While button clicks were the easiest, voice recognition introduced slightly more load as it required focus on spoken instructions or feedback. The Point and Select interface added further load, as it required temporarily removing a hand from the wheel and diverting visual attention. P4 noted, "When selecting objects by pointing with a finger, I should use my hand for a while. But it was not so burdensome since it finished very quickly, so I think it's not more difficult than speech recognition." Participants found touchscreen and text input particularly challenging. P6 commented, "It doesn't make sense to input by keyboard or just drag with a cursor while driving," and added, "When using navigation through speech recognition, I need to know the name of the place and pronounce it well. It is usually hard. If I can simply point and select like the Point and Select, it becomes incomparably easier."
\section{Discussion}
\subsection{Preserving Safety Margins through Cognitive Pacing}

By evaluating Point and Select under the strict visual and cognitive constraints of SAE Level 2 manual driving, we established a worst-case safety baseline. Our results demonstrate that the temporal decoupling approach successfully minimizes interference with primary vehicle control. The interaction proved highly efficient, achieving an average accuracy of 96\% and a task completion time of 1.82 seconds across urban speeds ranging from 30 to 70 km/h (Figure \ref{fig:SR}). This is notably faster than conventional IVIS tasks, which typically require over 2 seconds just to read information and 10--20 seconds for simple searches \cite{blanco2006impact}. 

\begin{figure}[t]
  \centering
  \includegraphics[width=0.8\columnwidth]{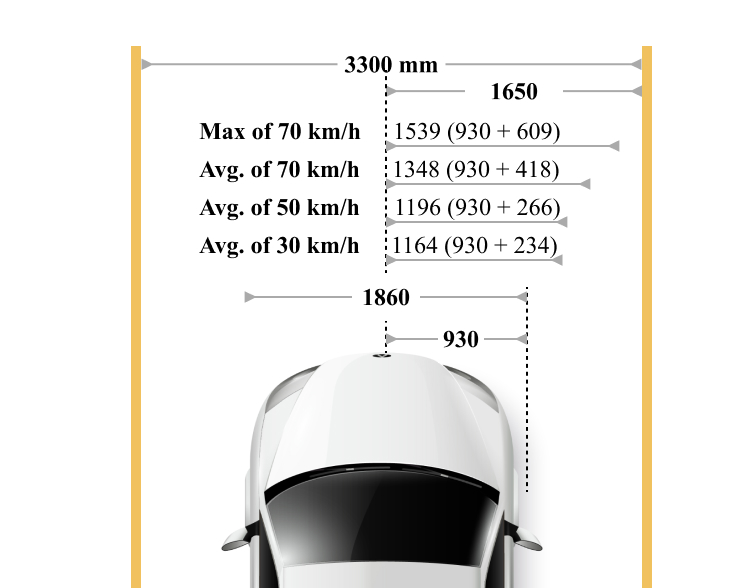}
  \caption{Analysis of lateral spatial occupancy within standard Korean urban lanes based on LLM data}
  \label{fig:occupy}
\end{figure}

Despite drivers completing at least 25 additional selections during a 5-minute drive, the impact on primary driving performance remained strictly within legal safety limits. Although statistical significance was observed for LLM at 70 km/h and SM at 50 km/h, the absolute deviations were practically negligible. For lateral control, the largest average deviation from the lane center was 418 mm at 70 km/h. Even the most extreme individual deviation (609 mm at 70 km/h by P3) was safely bounded within the 1650 mm margin of standard 3300 mm Korean urban lanes (Figure \ref{fig:occupy}). Similarly, for longitudinal control, the maximum speed deviation was 3.16 km/h at 50 km/h. This strictly adheres to the permissible error margin defined by Korean Automotive Safety Standards ($0 \leq \text{error} \leq \frac{\text{actual speed}}{10} + 6 \, \text{km/h}$), which allows deviations of up to 9, 11, and 13 km/h at speeds of 30, 50, and 70 km/h, respectively. Ultimately, proving its stability in this highly demanding Level 2 environment implies that as vehicles progress toward higher levels of automation, the safety margin of this interaction will be even further guaranteed.

Furthermore, our experimental setup required frequent, system-prompted interactions to rigorously evaluate the system under high workload. In everyday driving, external referencing is typically self-paced, allowing drivers to defer such tasks during demanding situations. Additionally, the limited field of view in simulators can negatively impact driving performance \cite{parduzi2019evaluation}. While a broader natural field of view and familiarity with actual vehicle dynamics could mitigate the observed impacts, though real-world traffic variables necessitate careful validation.

\subsection{Additional Considerations for Application in Real Road Environment}

While our simulation established a rigorous baseline for lateral and longitudinal control, it inherently excluded dynamic real-world actors, such as surrounding traffic and pedestrians, to minimize confounding variables. In actual urban driving—particularly in residential areas with ambiguous boundaries—the sudden appearance of vulnerable road users demands constant anticipation and readiness to decelerate \cite{braunagel2017ready}. These unpredictable events induce sudden spikes in cognitive workload. However, this reality underscores the critical advantage of our temporal decoupling design: if a sudden hazard arises, drivers can instantly abandon the interaction to attend to the road without losing their input. They can safely resume the deferred fine-selection phase via the steering wheel once the cognitive spike subsides.

Furthermore, the safe deployment of in-vehicle spatial interactions can be significantly augmented by integration with Advanced Driver Assistance Systems (ADAS). Technologies such as forward collision warning \cite{Nazan2016} and lane departure warning \cite{Liu2008} not only mitigate driving deviations \cite{blanco2006impact} but can also serve as a foundational safety net for secondary tasks. Building on this concept, future iterations of Point and Select could employ context-aware interaction management \cite{mok2017tunneled}. For instance, if ADAS detects a complex traffic pattern, crossing pedestrians, or high-speed cornering, the system could dynamically simplify the visual feedback or temporarily lock the fine-selection interface to prevent cognitive overload.

\subsection{Multimodal Enhancements and Collaborative Interactions}
While Point and Select relies on manual input, integrating other modalities can resolve the inherent ambiguities of spatial pointing \cite{gomaa2020studying, weidner2019interact}. For instance, eye-gaze tracking can refine the initial rough pointing, reducing the cognitive effort needed for fine selection. Concurrently, voice commands can provide semantic context, distinguishing between a navigational command (e.g., ``Guide me there'') and an informational query (e.g., ``What is that building?''). Furthermore, since our observations showed that referencing frequently occurs during passenger conversations, this interaction can naturally extend into a collaborative experience. The driver can perform the rapid rough pointing, while a passenger completes the fine selection on a shared display, preserving driving safety while fostering social engagement.

\subsection{Resolving Contextual Ambiguity in Real-World Applications}
Real-world application requires interpreting the specific intent behind a pointing gesture, as a single spatial indication can convey diverse meanings depending on the situational context. Future implementations can leverage context-aware systems or Large Language Mode-based intelligent agents to infer these layered intents. Additionally, extending the target scope from static buildings to dynamic infrastructure—such as traffic signs or parking systems—could unlock broader utilities \cite{chang2017eyes, yu2020social, merlino2019crossing}. However, enabling interactions with safety-critical targets requires cautious design to ensure the primary driving task is never compromised.

\subsection{Limitations and Future Work}
This study is inherently a possesses three main limitations. First, the current prototype requires an initial calibration per user to map individual pointing kinematics. Future iterations could leverage in-cabin driver monitoring systems (DMS) for automatic, zero-effort calibration. Second, our participants were mostly experienced drivers. Subsequent studies must evaluate novice or older drivers, as their cognitive thresholds and biomechanical responses may differ. Finally, to establish a controlled worst-case baseline for manual driving, our simulation deliberately excluded dynamic elements like surrounding traffic and unpredictable pedestrians. Real-world on-road validation is essential to understand how these external stressors affect a driver's interaction behaviors in everyday urban environments.
\section{Conclusion}
In this study, we introduced Point and Select, a spatial interaction technique that enables drivers to intuitively select external objects into the IVIS. Our video ethnography revealed that drivers naturally employ approximate pointing and actively defer complex interactions to manage cognitive load in highly dynamic driving contexts. Based on these insights, we designed a temporally decoupled system consisting of two phases: an initial rough pointing phase to rapidly capture a spatial target, followed by a deferred fine selection phase operated via steering-wheel buttons to mitigate cognitive friction by aligning with the driver’s natural behaviors.

A user study in a simulated SAE Level 2 manual driving environment demonstrated the system's robust efficiency, achieving an average success rate of 96.9\% and a task completion time of 1.82 seconds across urban speeds (30-70 km/h). despite the high visual and cognitive demands of manual driving, Point and Select maintained primary vehicle control—both lateral lane positioning and speed maintenance—within legal safety bounds, successfully establishing a foundational safety baseline for spatial referencing in high-demand, non-autonomous driving environments. Finally, we discuss the applicability of Point and Select in real-world driving contexts and explore its future potential in advanced automotive systems.

\section{Conflicts of interest}
The authors declare that they have no competing interests.

\section{Data availability}
The data underlying this article cannot be shared publicly due to the privacy of the individuals who participated in the study and ethical restrictions. The dataset includes identifiable video recordings and personal interview transcripts, which are strictly confidential in accordance with the Institutional Review Board (IRB) guidelines.

\section{Author contributions statement}
[Author contributions are blinded for peer review.]

\section{Acknowledgments}
[Acknowledgments are blinded for peer review.]

\bibliographystyle{oup-plain}
\bibliography{references}






\end{document}